\documentclass[aps,pra,reprint,groupedaddress]{revtex4-1}
\usepackage{graphicx}
\usepackage{color}
\usepackage{bm}
\usepackage{comment}
\usepackage{amssymb, amsmath}

\begin{document}


\title{Fast elementary gates for universal quantum computation with Kerr parametric oscillator qubits}


\author{Taro Kanao}
\email[]{taro.kanao@toshiba.co.jp}
\author{Hayato Goto}
\affiliation{Frontier Research Laboratory, Corporate Research \& Development Center, Toshiba Corporation, 1, Komukai-Toshiba-cho, Saiwai-ku, Kawasaki 212-8582, Japan}

\date{\today}

\begin{abstract}
Kerr parametric oscillators (KPOs) can stabilize the superpositions of coherent states, which can be utilized as qubits, and are promising candidates for realizing hardware-efficient quantum computers.
Although elementary gates for universal quantum computation with KPO qubits have been proposed, these gates are usually based on adiabatic operations and thus need long gate times, which result in errors caused by photon loss in KPOs realized by, e.g., superconducting circuits.
In this work, we accelerate the elementary gates by experimentally feasible control methods, which are based on numerical optimization of pulse shapes for shortcuts to adiabaticity.
By numerical simulations, we show that the proposed methods can achieve speedups compared to adiabatic ones by up to six times with high gate fidelities of 99.9\%.
These methods are thus expected to be useful for quantum computers with KPOs.
\end{abstract}


\maketitle

\section{Introduction}\label{sec_intro}
Towards hardware-efficient quantum computing, qubits with stabilized coherent states have been proposed~\cite{Cochrane1999, Mirrahimi2014}.
Coherent states with opposite phases and their superposition so-called a Schr\"{o}dinger cat state~\cite{Yurke1986, Deleglise2008} can be stabilized in a parametric oscillator with engineered two-photon dissipation~\cite{Wolinsky1988, Mirrahimi2014} or Kerr nonlinearity~\cite{Wielinga1993, Cochrane1999, Goto2016, Goto2016a, Puri2017a, Goto2019a}, which are referred to as a dissipative-cat qubit or a Kerr-cat qubit, respectively.
The latter is also called a Kerr parametric oscillator (KPO) qubit~\cite{Goto2016, Nigg2017, Wang2019, Xu2022, Yamaji2022, Masuda2022a, Kwon2022}.
In these qubits, when the two coherent states are used as computational basis, bit-flip errors can be suppressed, because a coherent state is robust against photon loss~\cite{Mirrahimi2014, Cochrane1999, Puri2017a}.
The stabilization of the coherent states and the suppression of bit-flip errors have been experimentally realized in superconducting circuits for the dissipative-cat qubit~\cite{Leghtas2015, Lescanne2020} and the KPO qubit~\cite{Wang2019, Grimm2020}.

A KPO does not rely on dissipation and can be described by a simple Hamiltonian.
Despite the simplicity, KPOs yield rich nonlinear dynamics such as quantum bifurcation~\cite{Cochrane1999, Goto2016, Puri2017a} and chaos~\cite{Milburn1990, Milburn1991, Goto2021b}.
The quantum bifurcation can be applied to quantum annealing~\cite{Goto2016} and a number of its implementations have been proposed~\cite{Nigg2017, Puri2017, Zhao2018, Goto2018a, Onodera2020, Goto2020b, Kanao2021, Yamaji2022a}.
By regarding two branches of the bifurcation as up and down spin states, a KPO lattice can behave like an Ising model, and its physics such as phase transitions has been studied~\cite{Savona2017, Dykman2018, Rota2019, Rota2019a, Kewming2020, Verstraelen2020, Miyazaki2022a}.
Other theoretical results on KPOs have been reported, such as exact solutions~\cite{Bartolo2016, Roberts2020}, state generations~\cite{Zhang2017, Goto2019b, Teh2020, Xue2022, Suzuki2023}, measurements and outputs~\cite{Bartolo2017, Masuda2021b, Matsumoto2023, Strandberg2021}, excited-state quantum phase transitions~\cite{Wang2020, Chavez-Carlos2023}, controls under a strong pump field~\cite{Masuda2021}, engineered dissipation~\cite{Putterman2022, Gautier2022, Ruiz2023}, a four-photon KPO~\cite{Kwon2022}, symmetries~\cite{Iachello2023}, and Floquet theory~\cite{Garcia-Mata2023}.

Applications of KPOs to fault-tolerant quantum computing~\cite{Nielsen2000} have also been studied~\cite{Puri2019}.
Quantum gates preserving the bias of errors mentioned above have been proposed~\cite{Puri2020}, which can be utilized for hardware-efficient quantum error correction~\cite{Darmawan2021a}.
Analytically engineered control methods for shortening the gate times of the bias-preserving gates have recently been proposed~\cite{Xu2022}.
Furthermore, for noisy intermediate-scale quantum (NISQ) applications~\cite{Preskill2018}, variational quantum algorithms~\cite{Cerezo2020, Endo2021} for KPOs have been proposed, such as quantum supervised machine learning~\cite{Mori2023} and a quantum approximate optimization algorithm~\cite{Vikstal2023}.

For implementing a KPO with a superconducting circuit, a Josephson parametric oscillator~\cite{Yamamoto2008, Wilson2010, Lin2014} with low photon loss has been suggested~\cite{Goto2016, Puri2017a}, and demonstrated experimentally~\cite{Wang2019}.
Then, by using a KPO in a three-dimensional cavity, single-qubit gates have been performed~\cite{Grimm2020}.
Also, tunable coupling between two KPOs has been realized~\cite{Yamaji2023}.
Other experiments with KPOs have been reported, such as a crossover from a Duffing oscillator to a KPO~\cite{Yamaji2022}, degenerate excited states~\cite{Frattini2022, Venkatraman2022}, single-qubit operations and characterizations with an ancillary transmon~\cite{Iyama2023}, and reflection coefficient measurements~\cite{Yamaguchi2023}.

For KPO qubits, elementary gates for universal quantum computation have been proposed~\cite{Goto2016a, Puri2017a}, which are based on adiabatic evolution and consist of $Z, X$, and $ZZ$ rotations denoted by $R_z, R_x$, and $R_{zz}$, respectively.
Experimentally, a study~\cite{Grimm2020} has demonstrated adiabatic $R_z$ and nonadiabatic $R_x$, and another study~\cite{Iyama2023} has adiabatically performed both $R_z$ and $R_x$.
Theoretically, other kinds of gate implementations have been proposed~\cite{Kanao2022b, Masuda2022a, Chono2022a, Aoki2023, Kang2022, Kang2023}.

Shorter gate times are desirable, because they can reduce errors caused by photon loss in KPOs and also enable faster computation.
However, the previous adiabatic elementary gates~\cite{Goto2016a, Puri2017a} need long gate times and otherwise diabatic transitions out of a qubit space cause leakage errors.
To reduce leakage errors, in this work, we focus on control methods called shortcuts to adiabaticity (STA)~\cite{Guery-Odelin2019}.
For KPOs, STA have been proposed for cat-state generation~\cite{Puri2017a, Goto2019b} and $R_{zz}$ with a phase rotation of a parametric drive~\cite{Masuda2022a}.
Also, a variant of the derivative removal by adiabatic gate (DRAG) technique, which is related to STA, has been proposed for the bias-preserving gates~\cite{Xu2022}.

To accelerate the elementary gates for universal quantum computation with KPO qubits, our approach is based on an STA called counterdiabatic terms (or counter terms for short)~\cite{Demirplak2003, Berry2009}, but does not use the exact counter terms, which are often experimentally infeasible.
Instead, we first approximate the counter terms by experimentally feasible terms~\cite{Opatrny2014} and then numerically optimize the pulse shapes for the gate operations.
As a result, we successfully shorten gate times, keeping high gate fidelities.
By this approach, the gate operations become faster by 2.6 times for $R_z$, 6.0 times for $R_{zz}$, and 2.6 times or higher for $R_x$.
Interestingly, the states of KPOs during the optimized gate operations for the shortest gate times are not necessarily instantaneous eigenstates, which indicates that the numerical optimization explores gate operations beyond the STA.
We also numerically show that the optimized gate operations are robust against systematic errors in the amplitudes of gate pulses, and the shortened gate times can suppress errors caused by single-photon loss.
We expect that these optimized elementary gates for KPO qubits will be useful for NISQ applications in a near term and fault-tolerant quantum computation in a long term.

\section{Approximate STA}\label{sec_method}
\subsection{Elementary gates for KPO qubits}
We first introduce the model of the KPO and elementary gates for universal quantum computation with the KPO qubits~\cite{Goto2016, Puri2017a}.
In a rotating frame and within the rotating-wave approximation, the Hamiltonian for a KPO is given by~\cite{Wielinga1993}
\begin{eqnarray}
	H_{\rm KPO}&=&-\frac{K}{2}a^{\dagger2}a^2+\frac{p}{2}\left(a^{\dagger2}+a^2\right),\label{eq_Hkpo}
\end{eqnarray}
where $a, K$, and $p$ are the annihilation operator, the Kerr coefficient, and the amplitude of a parametric drive, respectively.
In this study, the reduced Plank constant $\hbar$ is set to $1$.
The two degenerate eigenstates of the Hamiltonian corresponding to effective ground states of the KPO~\cite{Kanao2022b} are written as

\begin{eqnarray}
	|C_\pm\rangle&=&\frac{1}{\sqrt{2\left(1\pm e^{-2\alpha^2}\right)}}\left(|\alpha\rangle\pm|\!-\!\alpha\rangle\right),
\end{eqnarray}
where $|\!\pm\!\alpha\rangle$ are coherent states with an amplitude $\alpha=\sqrt{p/K}$.
In this work, we use the following computational basis~\cite{Puri2020, Kanao2022b},
\begin{eqnarray}
	|\tilde{0}\rangle&=&\frac{1}{\sqrt{2}}\left(|C_+\rangle+|C_-\rangle\right),\label{eq_til0}\\
	|\tilde{1}\rangle&=&\frac{1}{\sqrt{2}}\left(|C_+\rangle-|C_-\rangle\right),\label{eq_til1}
\end{eqnarray}
which are exactly orthogonal.
Equations~(\ref{eq_til0}) and (\ref{eq_til1}) are approximately equal to $|\!\pm\!\alpha\rangle$, respectively, for $p/K=4$ used in this study.

For the KPO qubits, elementary gates for universal computation can consist of $Z, X$, and $ZZ$ rotations, which are expressed respectively as~\cite{Nielsen2000}
\begin{eqnarray}
	R_z(\phi)&=&\left(\begin{array}{cc}
		e^{-i\phi/2}&0\\
		0&e^{i\phi/2}
	\end{array}\right),\\
	R_x(\theta)&=&\left(\begin{array}{cc}
		\cos\frac{\theta}{2}&-i\sin\frac{\theta}{2}\\
		-i\sin\frac{\theta}{2}&\cos\frac{\theta}{2}
	\end{array}\right),\\
	R_{zz}(\Theta)&=&\left(\begin{array}{cccc}
		e^{-i\Theta/2}&0&0&0\\
		0&e^{i\Theta/2}&0&0\\
		0&0&e^{i\Theta/2}&0\\
		0&0&0&e^{-i\Theta/2}
	\end{array}\right),
\end{eqnarray}
where $\phi, \theta$, and $\Theta$ are respective rotation angles.
For universal computation, arbitrary $\phi$, $\theta=\pi/2$, and $\Theta=\pi/2$ are enough~\cite{Goto2016a, Puri2017a, Nielsen2000}.
For KPOs, these elementary gates can be implemented based on adiabatic control with a single-photon drive, a detuning, and a linear coupling, respectively.
The Hamiltonians corresponding to the single-qubit gates are
\begin{eqnarray}
	H_0(t)&=&H_{\rm KPO}+g_0(t)A_0,\label{eq_H01bit}\\
	A_0&=&a^\dagger+a,\text{ for $R_z$},\label{eq_SingleDrive}\\
	A_0&=&a^\dagger a,\text{ for $R_x$},\label{eq_Detune}
\end{eqnarray}
where $g_0(t)$ is the amplitude of a gate pulse.
A linear coupling necessary for $R_{zz}$ can be realized with beam-splitter coupling~\cite{Goto2016a, Puri2017a} or two-mode squeezing~\cite{Xu2022}, described by
\begin{eqnarray}
	\!H_0(t)\!&=&\!H_{{\rm KPO}1}+H_{{\rm KPO}2}+\!g_0(t)A_0,\label{eq_H02bit}\\
	A_0\!&=&\!a_1^\dagger a_2\!+\!a_2^\dagger a_1,\text{for beam-splitter coupling},\label{eq_BeamSplitter}\\
	A_0\!&=&\!a_1^\dagger a_2^\dagger\!+\!a_1a_2,\text{for two-mode squeezing},\label{eq_TwoMode}
\end{eqnarray}
where $a_i$ and $H_{{\rm KPO}i}$ are the annihilation operator and the Hamiltonian in Eq.~(\ref{eq_Hkpo}) for the $i$th KPO.

\subsection{Approximate counter terms for STA}\label{sec_ApproxCounter}
An ideal counter term $H_1(t)$ for STA exactly reproduces adiabatic evolution with $H_0(t)$ by finite-time evolution with $H_0(t)+H_1(t)$~\cite{Guery-Odelin2019, Demirplak2003, Berry2009}, but is often experimentally infeasible.
Under certain assumptions, we approximate $H_1(t)$ by
\begin{eqnarray}
	H_1(t)&\simeq&\frac{i\dot{g}_0(t)}{4K\alpha^2}\left(a^\dagger-a\right),\text{ for $R_z$},\label{eq_cdz}\\
	H_1(t)&\simeq&\frac{i\dot{g}_0(t)}{4K\alpha^2}\left(a^{\dagger2}-a^2\right),\text{ for $R_z$},\label{eq_cdx}\\
	H_1(t)&\simeq&\frac{i\dot{g}_0(t)}{2K\alpha^2}\left(a_1^\dagger a_2^\dagger-a_1a_2\right),\text{ for $R_{zz}$},\label{eq_cdzz2}
\end{eqnarray}
where the dots denote the time derivative (See Appendix~\ref{sec_ApproxCD} for the details of the assumptions and derivations).
Importantly, these $H_1(t)$ are experimentally feasible as follows.
\begin{itemize}
\item $H_1(t)$ for $R_z$ in Eq.~(\ref{eq_cdz}) can be implemented with a single-photon drive with its phase shifted by $\pi/2$ from the original single-photon drive in Eq.~(\ref{eq_SingleDrive}).
\item $H_1(t)$ for $R_x$ in Eq.~(\ref{eq_cdx}) can be realized by a two-photon drive with its phase shifted by $\pi/2$ from the original parametric drive.
\item $H_1(t)$ for $R_{zz}$ in Eq.~(\ref{eq_cdzz2}) is another two-mode squeezing than the original one in Eq.~(\ref{eq_TwoMode}), which can be realized in a previously proposed superconducting circuit for $R_{zz}$~\cite{Chono2022a}.
\end{itemize}

Note that the counter term in Eq.~(\ref{eq_cdzz2}) can be derived from both $R_{zz}$ by the beam-splitter coupling in Eq.~(\ref{eq_BeamSplitter}) and $R_{zz}$ by the two-mode squeezing in Eq.~(\ref{eq_TwoMode}).
However, we numerically find that the two-mode squeezing in Eq.~(\ref{eq_TwoMode}) gives better results with the counter term in Eq.~(\ref{eq_cdzz2}), which can be understood from the matrix elements of $A_0$ and $H_1(t)$ (see Appendix~\ref{sec_Counter} for details).
We thus use the two-mode squeezing in Eq.~(\ref{eq_TwoMode}) in the following.

Here we also comment on another candidate of a counter term for $R_{zz}$,
\begin{eqnarray}
	H_1(t)\propto i\left(a_1^\dagger a_2-a_2^\dagger a_1\right).\label{eq_BeamSplitterOrth}
\end{eqnarray}
We numerically found that this term does not work as a counter term (a similar result has been mentioned in Ref.~\cite{Xu2022}).
Equation~(\ref{eq_BeamSplitterOrth}) does not cancel unwanted transitions out of the qubit space, because $H_1(t)$ in Eq.~(\ref{eq_BeamSplitterOrth}) and $A_0$ in Eqs.~(\ref{eq_BeamSplitter}) and (\ref{eq_TwoMode}) have different permutation symmetry, namely, symmetry with respect to the interchange of KPO1 and 2 (see Appendix~\ref{sec_Counter} for details).

\subsection{Numerical optimization}
To go beyond the analytic approximate $H_1(t)$ in Eqs.~(\ref{eq_cdz})-(\ref{eq_cdzz2}), our proposed approach uses arbitrary waveforms for the amplitudes of the counter pulses $g_1(t)$ as
\begin{eqnarray}
	H_1(t)&=&g_1(t)A_1,\label{eq_HCounter}\\
	A_1&=&i\left(a^\dagger-a\right),\text{ for $R_z$},\label{eq_A1Rz}\\
	A_1&=&i\left(a^{\dagger2}-a^2\right),\text{ for $R_x$},\\
	A_1&=&i\left(a_1^\dagger a_2^\dagger-a_1a_2\right),\text{ for $R_{zz}$},
\end{eqnarray}
and numerically optimizes $g_1(t)$ as well as $g_0(t)$ in Eqs.~(\ref{eq_H01bit}) and (\ref{eq_H02bit}).
Total Hamiltonians are then given by, for the single- and two-qubit gates, respectively,
\begin{eqnarray}
	H(t)&=&H_{\rm KPO}+g_0(t)A_0+g_1(t)A_1,\label{eq_Ht1}\\
	H(t)&=&H_{{\rm KPO}1}+H_{{\rm KPO}2}+g_0(t)A_0+g_1(t)A_1,\label{eq_Ht2}
\end{eqnarray}
where $A_0$ are given in Eqs.~(\ref{eq_SingleDrive}), (\ref{eq_Detune}), and (\ref{eq_TwoMode}).
Here, the two-mode squeezing Hamiltonian in Eq.~(\ref{eq_TwoMode}) is used as mentioned above.
We expect that this approach, which numerically optimizes pulse shapes for STA, will be useful for other qubit systems.

To optimize $g_0(t)$ and $g_1(t)$ numerically, we express the waveforms of the pulse amplitudes by~\cite{Martinis2014}
\begin{eqnarray}
	g_0(t)&=&\sum_{n=1}^{N_f}\bigg[g_{0,2n-1}\sin\frac{(2n-1)\pi t}{T}\nonumber\\
		&&+\frac{g_{0,2n}}{2}\left(1-\cos\frac{2\pi nt}{T}\right)\bigg],\label{eq_g0}\\
	g_1(t)&=&\sum_{n=1}^{N_f}g_{1,n}\sin\frac{2\pi nt}{T},\label{eq_g1}
\end{eqnarray}
where $T$ is a gate time and $N_f$ determines the number of frequency components.
We choose the symmetric $g_0(t)$ and antisymmetric $g_1(t)$ with respect to time reversal $t\to T-t$, because an exact counter term is antisymmetric when the other term is symmetric (see Appendix~\ref{sec_TimeReversal}).
In $g_0(t)$, we include the sine terms to allow for nonzero $\dot{g}_0(t)$ at $t=0, T$~\cite{Goto2016a}.
Since the highest frequencies in $g_j(t)$ are limited to $N_f/T$ and $g_j(t)$ are zero at initial and final times, these waveforms are expected to be experimentally feasible.

We numerically optimize $g_{j,n}$ in Eqs~(\ref{eq_g0}) and (\ref{eq_g1}) to maximize an average gate fidelity $\bar{F}$~\cite{Nielsen2002, Pedersen2007} given in Eq.~(\ref{eq_BarF}) in Appendix~\ref{sec_AverageFidelity}, using the quasi-Newton method with the BFGS formula~\cite{MATLAB}.
We set the initial values of $g_{j,n}$ for the optimization to the ones corresponding to analytic waveforms for adiabatic elementally gates without and with the counter terms in Eqs.~(\ref{eq_cdz})-(\ref{eq_cdzz2}) (see Appendix~\ref{sec_wave}).

\section{Numerical simulations}\label{sec_result}
In the present simulations, we regard the Kerr coefficient $K$ as the unit of the frequency and set the amplitude of the parametric drive to $p=4K$.
We express states and operators in the photon-number basis with the largest photon number of 40, which is large enough.
We simulate the time evolution of states by numerically solving the Schr\"{o}dinger equation
\begin{eqnarray}
	i|\dot{\psi}\rangle&=&H(t)|\psi\rangle,\label{eq_Schr}
\end{eqnarray}
unless otherwise stated.
We use the fourth-order Runge-Kutta method with the step size of $10^{-4}/K$.

We compare the following four cases depending on the waveforms and the counter terms:
\begin{enumerate}
	\item Analytic waveforms without the counter terms.
	\item Analytic waveforms with the counter terms in Eqs.~(\ref{eq_cdz})-(\ref{eq_cdzz2}).
	\item Numerically optimized waveforms in Eq.~(\ref{eq_g0}) without the counter terms.
	\item Numerically optimized waveforms in Eqs.~(\ref{eq_g0}) and (\ref{eq_g1}) with the counter terms in Eq.~(\ref{eq_HCounter}).
\end{enumerate}
In this work $N_f$ in Eqs.~(\ref{eq_g0}) and (\ref{eq_g1}) is set to 10.
The analytic waveforms and initial values of $g_{j,n}$ for the numerical optimization are given in Appendix~\ref{sec_wave}.

\subsection{Simulation results for $R_z$}\label{sec_Rz}
\begin{figure}
	\includegraphics[width=8.2cm]{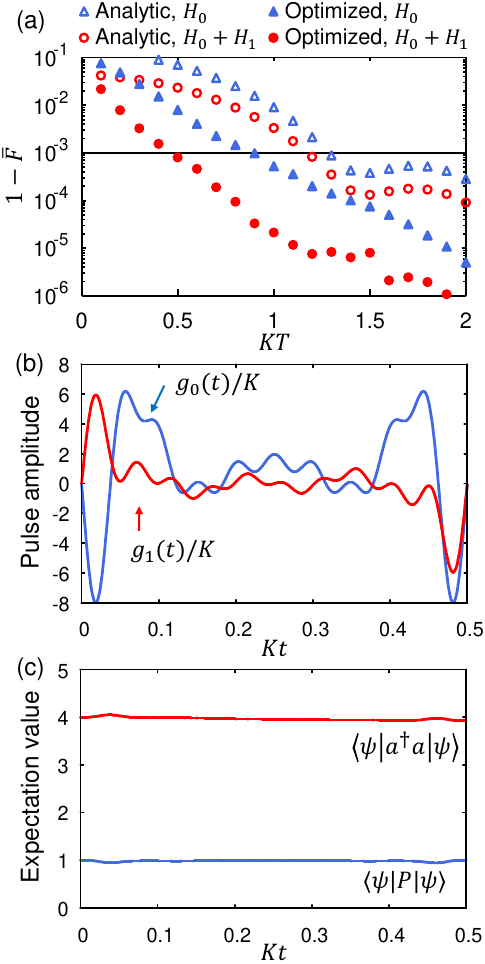}
	\caption{(a) Average infidelities for $R_z(\pi)$ as functions of gate time for an analytic waveform of a pulse amplitude without a counter term (Analytic, $H_1$) and with it (Analytic, $H_1+H_2$), and for numerically optimized waveforms without the counter term (Optimized, $H_1$) and with it (Optimized, $H_1+H_2$).
		The line indicates $1-\bar{F}=10^{-3}$.
		(b) Waveforms of the amplitudes of a gate pulse $g_0(t)$ and a counter pulse $g_1(t)$ for $R_z(\pi)$, which are numerically optimized for $KT_{\rm min}=0.5$.
		(c) Mean photon number and population in the qubit space during $R_z(\pi)$ with $g_j(t)$ in (b).
		The initial state is $|C_+\rangle$.
		\label{fig_ErT}}
\end{figure}
Figure~\ref{fig_ErT}(a) shows the average infidelities $1-\bar{F}$ for $R_z(\pi)$ as functions of the dimensionless gate time $KT$.
The infidelities decrease with increasing gate time, indicating the adiabaticity of the gate, where the errors are mainly due to the leakage of population to the states outside the qubit space.
We define a minimum gate time $T_{\rm min}$ by minimal $T$ satisfying $1-\bar{F}<10^{-3}$, and compare $T_{\rm min}$ for the above four cases.
With analytic waveforms, $KT_{\rm min}$ are $1.3$ and $1.2$, respectively without and with the counter term.
By the numerical optimization, $KT_{\rm min}$ are shortened to $0.9$ and $0.5$, respectively.
Thus the numerically optimized $R_z$ with the counter term is 2.6 times faster than the original analytic $R_z$ without the counter term.
These results show that the counter term is effective and the improvement is enhanced by the numerical optimization. 

We examine the optimized gate operation with the counter term at $KT_{\rm min}=0.5$.
The optimized waveforms of $g_j(t)$ are shown in Fig.~\ref{fig_ErT}(b).
Figure~\ref{fig_ErT}(c) shows the resulting time evolutions of the mean photon number and population in the qubit space with the initial state $|C_+\rangle$, where $P$ is a projector onto the computational basis states,
\begin{eqnarray}
	P=|\tilde{0}\rangle\langle\tilde{0}|+|\tilde{1}\rangle\langle\tilde{1}|.
\end{eqnarray}
It is notable that despite the large amplitudes of $g_j(t)$, the mean photon number and the population in the qubit space are almost unchanged.

To see the state in more detail, we use the Wigner function $W(x, y)$, which is a quasiprobability distribution for ${x=\left(a+a^\dagger\right)/2, y=\left(a-a^\dagger\right)/(2i)}$~\cite{Leonhardt1997} and is calculated by the technique in Ref.~\cite{Goto2016}.
\begin{figure}
	\includegraphics[width=8.2cm]{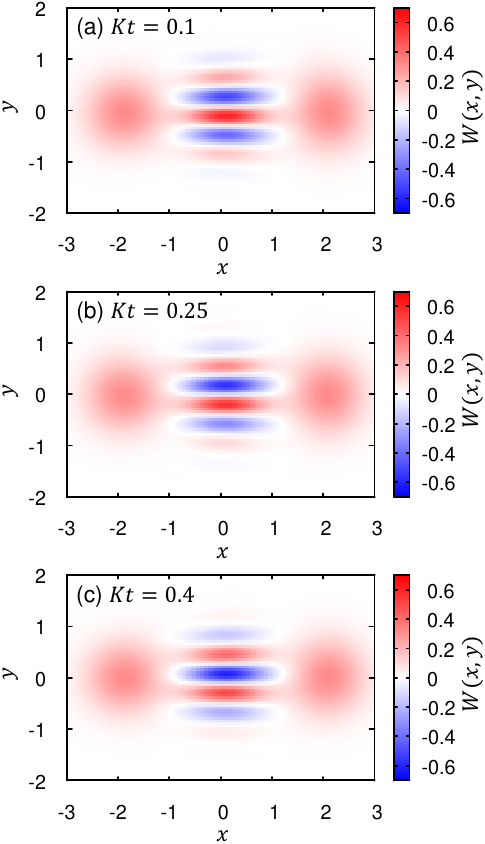}
	\caption{The Wigner functions during time evolution under the optimized $g_j(t)$ in Fig.~\ref{fig_ErT}(b).
		The initial state is $|C_+\rangle$.
		\label{fig_Wigner}}
\end{figure}
Figure~\ref{fig_Wigner} shows $W(x,y)$ during the gate operation with the optimized $g_j(t)$ in Fig.~\ref{fig_ErT}(b).
The gate operation retains the two peaks around ${x=\pm2, y=0}$ and the interference fringe between them, which indicate that the state is in the superposition of the coherent states.
Only the interference fringe changes with the time, corresponding to the relative phase rotations of $|\tilde{0}\rangle$ and $|\tilde{1}\rangle$.
These dynamics are possible because the single-photon drives used for $R_z$ can preserve the coherent states when the effective potential of the KPO is well approximated by the double well~\cite{Wang2019, Puri2019}.
Interestingly, we numerically found that the cat states in Fig.~\ref{fig_Wigner} are not instantaneous eigenstates of $H_0(t)$, which indicates that our proposed approach is beyond STA.

We next show that the optimized $g_j(t)$ with the counter term for $R_z(\pi)$ can be used for $R_z(\phi)$ with arbitrary $\phi$ by introducing only one time-independent scaling parameter $\lambda$.
The pulse amplitudes are set to $\lambda g_0(t)$ and $\lambda g_1(t)$.
Resulting $\phi$ is determined by maximizing $\bar{F}$.
\begin{figure}
	\includegraphics[width=8.2cm]{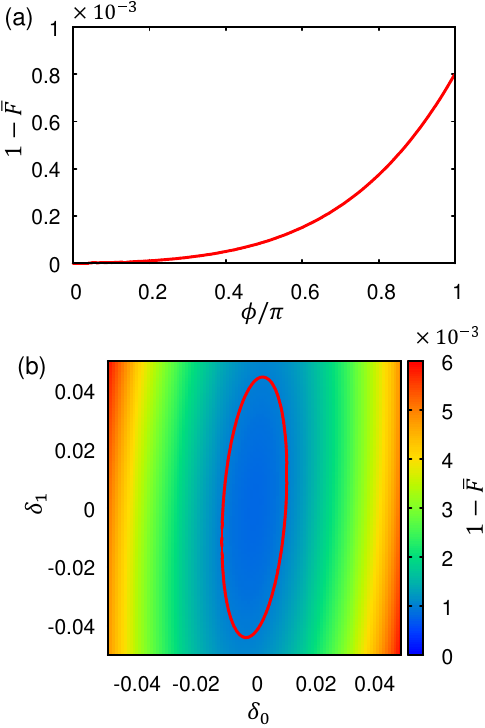}
	\caption{(a) Average infidelity for $R_z(\phi)$ obtained by $\lambda g_j(t)$ with $g_j(t)$ shown in Fig.~\ref{fig_ErT}(b), which are optimized to $R_z(\pi)$, and a time-independent scaling parameter $0\leq\lambda\leq1$.
		(b) Average infidelity for $R_z(\pi)$ obtained by $\left(1+\delta_j\right)g_j(t)$ with the optimized $g_j(t)$ and relative errors $\delta_j$.
		The line indicates $1-\bar{F}=10^{-3}$.
		\label{fig_ErPhi}}
\end{figure}
Figure~\ref{fig_ErPhi}(a) shows $1-\bar{F}$ as a function of $\phi$ at $KT_{\rm min}=0.5$, which demonstrates that this method gives high-fidelity $R_z(\phi)$ for arbitrary $\phi$ in $0\leq\phi\leq\pi$.
An exact counter term suggests that this continuous gate by the one parameter $\lambda$ is possible because the changes in the states are small during the gate operation as shown in Fig.~\ref{fig_Wigner} (see Appendix~\ref{sec_ContinuousGate} for details).
On the other hand, this continuous gate does not hold for $R_x(\theta)$ as also mentioned later in Sec.~\ref{sec_Rx}, where the states largely change during the gate operation.

To examine the optimality and robustness of the optimized $g_j(t)$, we evaluate $R_z(\pi)$ with ${\left(1+\delta_j\right)g_j(t)}$
for given relative errors $\delta_j$, which can model systematic errors in the pulse amplitudes~\cite{Masuda2022a}.
Figure~\ref{fig_ErPhi}(b) shows $1-\bar{F}$ as a function of $\delta_j$.
First, at $\delta_0=\delta_1=0$, the gradient of $1-\bar{F}$ with respect to $\delta_j$ vanishes, implying that 
$\delta_0=\delta_1=0$ is an optimal point.
Second, the ellipse in Fig.~\ref{fig_ErPhi}(b) shows the contour corresponding to $1-\bar{F}=10^{-3}$, indicating that such high-fidelity gate operation can be achieved even for the relative errors as large as $|\delta_0|=0.01$ or $|\delta_1|=0.05$.
In particular, this gate operation is robust for the error $\delta_1$, namely, the error in the counter pulse.

\subsection{Simulation results for $R_{zz}$}\label{sec_Rzz}
\begin{figure}
	\includegraphics[width=8.2cm]{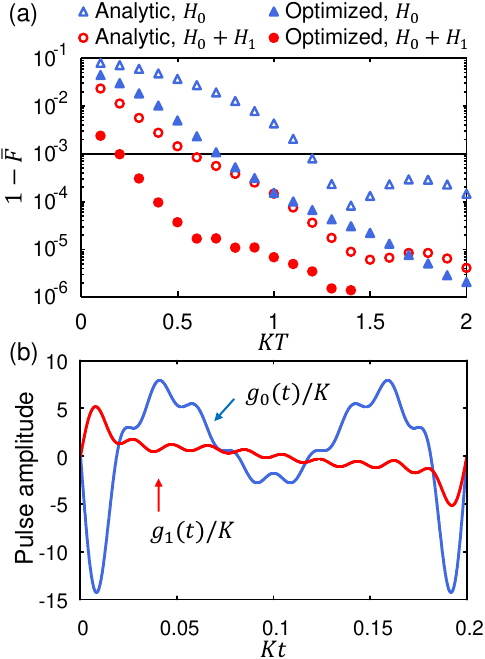}
	\caption{(a) Average infidelities for $R_{zz}(\pi/2)$ as functions of gate time.
		(b) Waveforms of $g_j(t)$ for $R_{zz}(\pi/2)$ with a counter term, optimized numerically for $KT_{\rm min}=0.2$.
		\label{fig_ErRzz}}
\end{figure}
Figure~\ref{fig_ErRzz}(a) shows $1-\bar{F}$ for $R_{zz}(\pi/2)$ as functions of $KT$.
With analytic waveforms, the minimum gate time satisfying $1-\bar{F}<10^{-3}$ is $KT_{\rm min}=1.2$ and $0.6$ without and with the counter term, respectively.
With the numerical optimization, corresponding $KT_{\rm min}$ are $0.8$ and $0.2$.
Thus, the numerically optimized gate operation with the counter term provides a speedup by 6.0 times compared with the original analytic waveform without the counter term.
The gate-time dependence of the infidelities of $R_{zz}$ in Fig.~\ref{fig_ErRzz}(a) is similar to that of $R_z$ in Fig.~\ref{fig_ErT}(a), which may be because for one KPO the other acts like a single-photon drive as in $R_z$.
\begin{figure}
	\includegraphics[width=8.2cm]{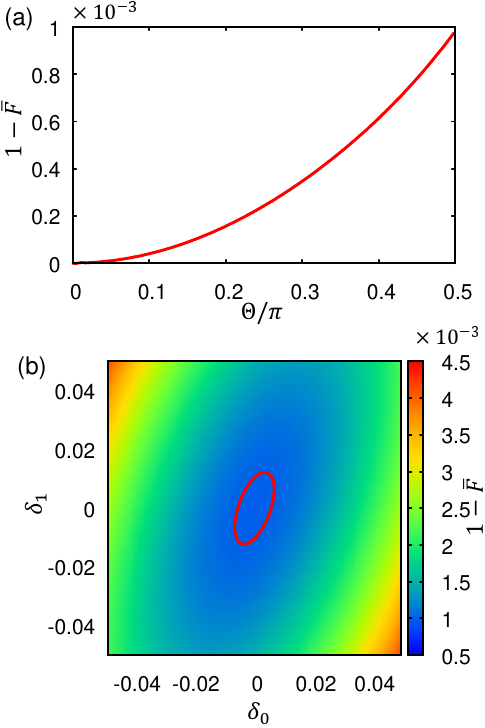}
	\caption{(a) Average infidelity for $R_{zz}(\Theta)$ obtained by $\lambda g_j(t)$ with the optimized $g_j(t)$ in Fig.~\ref{fig_ErRzz}(b).
		(b) Average infidelity for $R_{zz}(\pi/2)$ obtained by $(1+\delta_j)g_j(t)$ with the optimized $g_j(t)$.
		The line indicates $1-\bar{F}=10^{-3}$.
		\label{fig_WaveRzz}}
\end{figure}
Optimized waveforms $g_j(t)$ with the counter term at $KT_{\rm min}=0.2$ are shown in Fig.~\ref{fig_ErRzz}(b).
Figure~\ref{fig_WaveRzz}(a) shows that these optimized $g_j(t)$ can be used for continuous $R_{zz}(\Theta)$ by $\lambda g_j(t)$ with the time-independent scaling parameter $\lambda$ as in the case of $R_z(\phi)$.
Also, the optimality and robustness of $R_{zz}(\pi/2)$ are evaluated with $(1+\delta_j)g_j(t)$. Figure~\ref{fig_WaveRzz}(b) shows that the gradient of $1-\bar{F}$ is zero at $\delta_0=\delta_1=0$, indicating its optimality.

\subsection{Simulation results for $R_x$}\label{sec_Rx}
\begin{figure}
	\includegraphics[width=8.2cm]{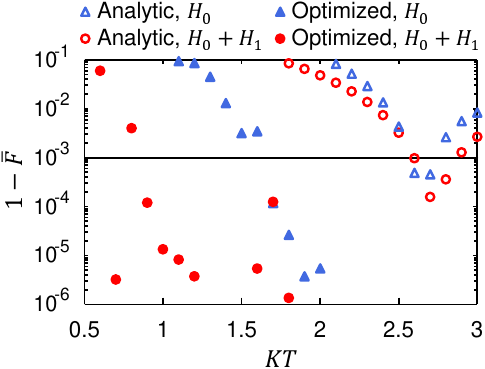}
	\caption{Average infidelities for $R_x(\pi/2)$ as functions of gate time.
		\label{fig_ErRx}}
\end{figure}
Figure~\ref{fig_ErRx} shows $1-\bar{F}$ for $R_x(\pi/2)$ as functions of $KT$.
With analytic waveforms, $KT_{\rm min}$ are $2.6$ both with and without the counter term.
With the numerical optimization, $KT_{\rm min}$ are $1.7$ and $0.6$ without and with the counter term, respectively, which means that our approach can achieve a 4.3 times faster gate operation than that with the original analytic waveform without the counter term.
However, we find that for $KT\leq0.9$, the maximum value of $|g_j(t)|/K$ can be larger than $50$, which might be infeasible because the rotating-wave approximation would be no longer valid~\cite{Masuda2021}.
\begin{figure}
	\includegraphics[width=8.2cm]{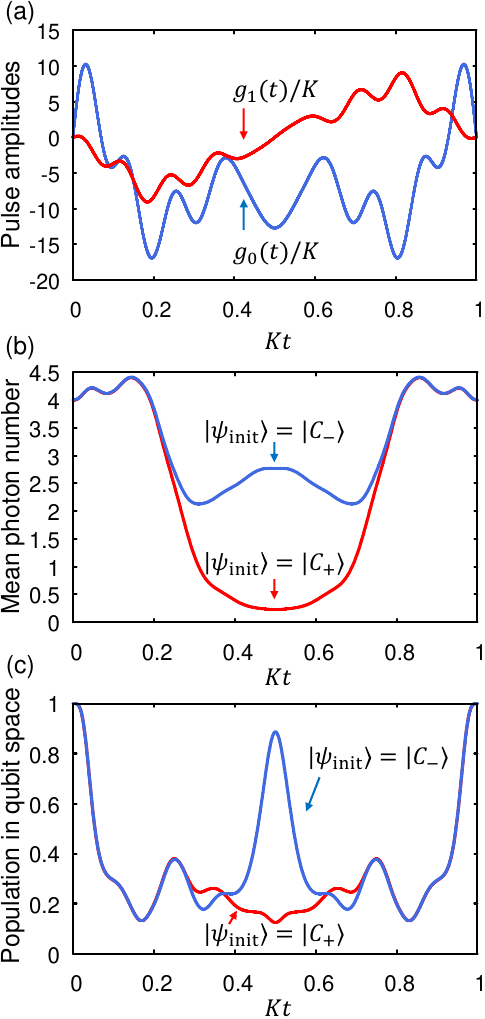}
	\caption{(a) Waveforms of the amplitudes of $g_j(t)$ for $R_x(\pi/2)$ with a counter term, optimized numerically for $KT_{\rm min}=1$.
		(b) Mean photon number and (c) population in the qubit space during the optimized $R_x(\pi/2)$.
		\label{fig_WaveRx}}
\end{figure}
Thus, in the following, we examine optimized $g_j(t)$ at $KT=1$, which is an acceleration by 2.6 times compared with the analytic waveform without the counter term.
Then, the pulse amplitudes with $|g_j(t)|/K<20$ are obtained as shown in Fig.~\ref{fig_WaveRx}(a).
Figures~\ref{fig_WaveRx}(b) and \ref{fig_WaveRx}(c) show the mean photon numbers $\langle\psi|a^\dagger a|\psi\rangle$ and the populations in the qubit space $\langle\psi|P|\psi\rangle$, respectively, during the optimized $R_x$.
These quantities become small during the operation, because the large detuning suppresses the oscillation of the KPO and alters the state.
(See Appendix~\ref{sec_WignerRx} for the Wigner function during the optimized $R_x$.)

As mentioned in Sec.~\ref{sec_Rz}, we find that $R_x(\theta)$ for continuous $\theta$ is not obtained by $\lambda g_j(t)$ with the optimized $g_j(t)$ for ${\theta=\pi/2}$.
This might be because the states largely change from the two coherent states during the optimized $R_x(\pi/2)$ unlike $R_z$ and $R_{zz}$ (see Appendix~\ref{sec_ContinuousGate}).

\begin{figure}
	\includegraphics[width=8.2cm]{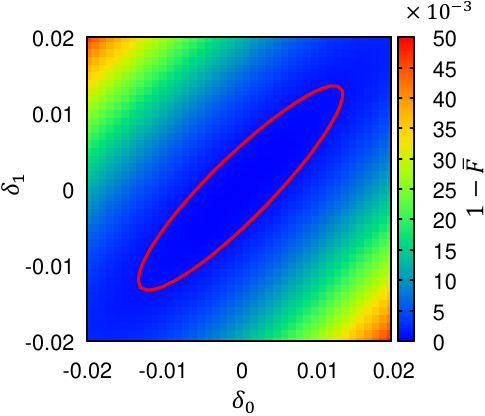}
	\caption{Average infidelity for $R_{x}(\pi/2)$ obtained by $(1+\delta_j)g_j(t)$ with the optimized $g_j(t)$ in Fig.~\ref{fig_WaveRx}(a).
		The line indicates $1-\bar{F}=10^{-3}$.
		\label{fig_ErDltRx}}
\end{figure}
The optimality and robustness of the optimized $g_j(t)$ are again evaluated by the gate operation by $\left(1+\delta_j\right)g_j(t)$.
Figure~\ref{fig_ErDltRx} shows the average infidelity as a function of $\delta_j$, indicating that the optimality and robustness hold.
Compared with $R_z$ and $R_{zz}$, the average infidelity for $R_x$ shows larger correlation between $\delta_0$ and $\delta_1$, that is, the optimized $R_x$ is more robust against the relative errors with $\delta_0\simeq\delta_1$.
This result suggests that the counter pulse $g_1(t)$ plays a more important role in $R_x$ than in the others.

\subsection{Effect of single-photon loss}
Finally, we evaluate errors in the presence of single-photon loss, which we choose as a representative of decoherence sources in KPOs.
We solve the master equation for a density operator $\rho$,
\begin{eqnarray}
	\dot{\rho}&=&-i[H(t),\rho]+\mathcal{L}[\rho],\label{eq_master}\\
	\mathcal{L}[\rho]&=&\frac{\kappa}{2}\left(2a\rho a^\dagger-a^\dagger a\rho-\rho a^\dagger a\right),\text{for $R_z, R_x$},\label{eq_master1}\\
	\mathcal{L}[\rho]&=&\!\frac{\kappa}{2}\!\sum_{i=1,2}\!\left(\!2a_i\rho a_i^\dagger\!-\!a_i^\dagger a_i\rho\!-\!\rho a_i^\dagger a_i\!\right)\!, \text{for $R_{zz}$},\label{eq_master2}
\end{eqnarray}
where $[O_1, O_2]=O_1O_2-O_2O_1$ is the commutation relation for operators and $\kappa$ is the loss rate.
$H(t)$ are given in Eqs.~(\ref{eq_Ht1}) and (\ref{eq_Ht2}), and the optimized $g_j(t)$ are used.
Here, we evaluate an average gate fidelity calculated with a finite number of initial states, $\bar{F}_{\rm loss}$, which is defined by Eq.~(\ref{eq_TilBarF}) in Appendix~\ref{sec_AverageFidelity}.

\begin{figure}
	\includegraphics[width=8.2cm]{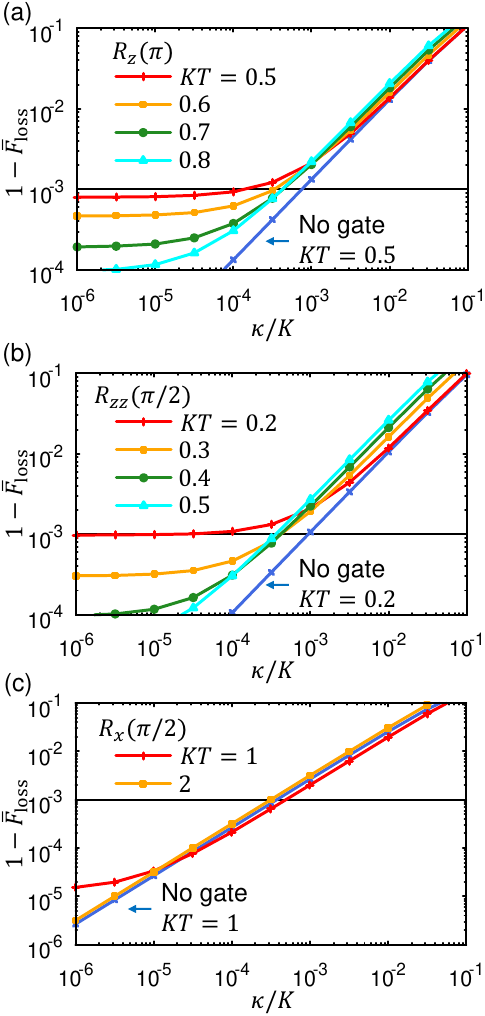}
	\caption{Average infidelities as functions of the loss rate for (a) $R_z(\pi)$, (b) $R_{zz}(\pi/2)$, and (c) $R_x(\pi/2)$, where optimized $g_j(t)$ are used.
		No gate means $g_j(t)=0$ and its ideal operation is no rotation.
		The horizontal line indicates $1-\bar{F}_{\rm loss}=10^{-3}$.
		\label{fig_ErKp}}
\end{figure}
Figure~\ref{fig_ErKp} shows $1-\bar{F}_{\rm loss}$ as functions of the loss rate.
For all of $R_z, R_{zz}$, and $R_x$, the errors can be suppressed below $1-\bar{F}_{\rm loss}<10^{-3}$ for the loss rate as large as $\kappa/K\leq3\times10^{-4}$, which can be achieved by optimizing the gate times $KT$ as follows.
For small $\kappa/K$, the errors are dominated by the leakage errors, which depend little on $\kappa/K$ and decrease with increasing $KT$.
For large $\kappa/K$, the errors are mainly due to dephasing caused by the single-photon loss~\cite{Puri2017a}, because $1-\bar{F}_{\rm loss}$ with and without the gate operations overlap for the same $KT$, especially for $R_z$ and $R_{zz}$ in Figs.~\ref{fig_ErKp}(a) and \ref{fig_ErKp}(b).
The dephasing errors increase with $KT$ [cf. Eq.~(\ref{eq_Floss}) in Appendix~\ref{sec_AverageFidelity}].
For intermediate $\kappa/K$, the leakage and the dephasing compete, and thus $1-\bar{F}_{\rm loss}$ are the smallest at larger $KT$ than $KT_{\rm min}$, achieving $1-\bar{F}_{\rm loss}<10^{-3}$ for $\kappa/K\leq3\times10^{-4}$ as above.

In Fig.~\ref{fig_ErKp}(c), for large $\kappa/K$, $R_x$ gives smaller $1-\bar{F}_{\rm loss}$ than no gate operation, because the mean photon number is decreased during $R_x$ as shown in Fig.~\ref{fig_WaveRx}(b) and the effect of single-photon loss becomes smaller [cf. Eq.~(\ref{eq_Floss}) in Appendix~\ref{sec_AverageFidelity}].

\section{Summary}\label{sec_summary}
We have shown that adiabatic elementary gates for universal qunatum computation with KPO qubits can be accelerated by utilizing feasible counter terms derived from STA and numerically optimizing the pulse shapes for them.
The optimized gate operations are feasible in experiments with, specifically, superconducting circuits.
We thus expect that the proposed methods are useful for quantum computers with KPOs.

\begin{acknowledgments}
We thank S. Masuda, Y. Matsuzaki, T. Ishikawa, S. Kawabata, and H. Chono for valuable discussion.
This paper is based on results obtained from a project, JPNP16007, commissioned by the New Energy and Industrial Technology Development Organization (NEDO), Japan.
\end{acknowledgments}

\appendix
\section{Approximate counter terms}\label{sec_ApproxCD}
We start from an exact counter term $H_1(t)$ given by~\cite{Berry2009, Guery-Odelin2019}
\begin{eqnarray}
	H_1(t)&=&i\sum_k\Big[|\dot{E}_k(t)\rangle\langle E_k(t)|\nonumber\\
	&&-\langle E_k(t)|\dot{E}_k(t)\rangle|E_k(t)\rangle\langle E_k(t)|\Big],\label{eq_H1}
\end{eqnarray}
where $|E_k(t)\rangle$ is the $k$th eigenstate of $H_0(t)$ with its eigenvalue $E_k(t)$,
\begin{eqnarray}
	H_0(t)|E_k(t)\rangle&=&E_k(t)|E_k(t)\rangle.\label{eq_eigen}
\end{eqnarray}
The first term in the right-hand side of Eq.~(\ref{eq_H1}) cancels transitions between different $|E_k(t)\rangle$, while the second term corrects phase factors.
We ignore the second term because we find that $\langle E_k(t)|\dot{E}_k(t)\rangle$ vanishes in the approximation below.
We then express $H_1(t)$ in an explicitly Hermitian form using the time derivative of $\sum_k|E_k(t)\rangle\langle E_k(t)|=I$ ($I$ is the identity operator) as
\begin{eqnarray}
	H_1(t)&\simeq&\frac{i}{2}\!\sum_k\!\left[|\dot{E}_k(t)\rangle\langle E_k(t)|\!-\!|E_k(t)\rangle\langle \dot{E}_k(t)|\right]\!.\label{eq_cd}
\end{eqnarray}
To cancel transitions from a qubit space, we restrict the summation to $k$ for computational basis states.

From Eq.~(\ref{eq_cd}), we derive approximate counter terms for KPOs.
We approximate $|E_k(t)\rangle$ corresponding to the qubit states by a variational method using a coherent state $|\beta\rangle$ as a trial state~\cite{Kanao2021}.
Its amplitude $\beta$ is determined by seeking an extremum with
\begin{eqnarray}
	\frac{\partial}{\partial\beta}\langle\beta|H_0(t)|\beta\rangle&=&0.\label{eq_variational}
\end{eqnarray}

\subsection{Approximate counter term for $R_z$}
For $R_z$, we approximately solve Eq.~(\ref{eq_variational}) by assuming $\left|g_0(t)/\left(2K\alpha^3\right)\right|\ll1$ and obtain the following expression for the qubit states,
\begin{eqnarray}
	|E_0(t)\rangle&\simeq&\left|\alpha+\frac{g_0(t)}{2K\alpha^2}\right\rangle,\\
	|E_1(t)\rangle&\simeq&\left|-\alpha+\frac{g_0(t)}{2K\alpha^2}\right\rangle,
\end{eqnarray}
where $\alpha=\sqrt{p/K}$.
For real $\beta(t)$,
\begin{eqnarray}
	|\dot{\beta}(t)\rangle=\dot{\beta}(t)\left[a^\dagger-\beta(t)\right]|\beta(t)\rangle,
\end{eqnarray}
holds.
Equation~(\ref{eq_cd}) can then give Eq.~(\ref{eq_cdz}) as
\begin{eqnarray}
	H_1(t)&\simeq&\frac{i\dot{g}_0(t)}{4K\alpha^2}\left[a^\dagger P(t)-P(t)a\right]\label{eq_cdzp}\\
	&\simeq&\frac{i\dot{g}_0(t)}{4K\alpha^2}\left(a^\dagger-a\right),\label{eq_cdzApp}
\end{eqnarray}
where $P(t)=|E_0(t)\rangle\langle E_0(t)|+|E_1(t)\rangle\langle E_1(t)|$ is a projector onto the qubit space.
$P(t)$ is ignored in Eq.~(\ref{eq_cdzApp}), because the difference between Eqs.~(\ref{eq_cdzp}) and (\ref{eq_cdzApp}) is proportional to $a^\dagger[I-P(t)]-[I-P(t)]a$, which has matrix elements mainly between states outside the qubit space, and therefore is negligible.

\subsection{Approximate counter term for $R_{zz}$}
The following derivation is valid for both the beam-splitter coupling in Eq.~(\ref{eq_BeamSplitter}) and the two-mode squeezing in Eq.~(\ref{eq_TwoMode}).
From Eq.~(\ref{eq_variational}), when $\left|g_0(t)/\left(2K\alpha^2\right)\right|\ll1$, the qubit states are approximately given by
\begin{eqnarray}
	\!|E_{s_1s_2}(t)\rangle&\simeq&\left|s_1\alpha\!+\!\frac{s_2g_0(t)}{2K\alpha}\right\rangle\!\left|s_2\alpha\!+\!\frac{s_1g_0(t)}{2K\alpha}\right\rangle\!,
\end{eqnarray}
where the states are labeled with $s_i=\pm1$ for $i=1,2$.
Equation~(\ref{eq_cd}) then becomes
\begin{widetext}
\begin{eqnarray}
	H_1(t)\simeq\frac{i\dot{g}_0(t)}{4K\alpha}\sum_{s_1=\pm1, s_2=\pm1}\left[\left(s_2a_1^\dagger+s_1a_2^\dagger\right)|E_{s_1s_2}(t)\rangle\langle E_{s_1s_2}(t)|-|E_{s_1s_2}(t)\rangle\langle E_{s_1s_2}(t)|\left(s_2a_1+s_1a_2\right)\right].\label{eq_cdzz1}
\end{eqnarray}
\end{widetext}
Using $\sum_{s_1s_2}s_i|E_{s_1s_2}(t)\rangle\langle E_{s_1s_2}(t)|\simeq P(t)a_iP(t)/\alpha=P(t)a_i^\dagger P(t)/\alpha$ with $P(t)=\sum_{s_1s_2}|E_{s_1s_2}(t)\rangle\langle E_{s_1s_2}(t)|$ and ignoring $P(t)$ in a similar manner to $R_z$, we obtain approximate $H_1(t)$ as
\begin{eqnarray}
	H_1(t)\simeq\frac{i\dot{g}_0(t)}{2K\alpha^2}\left(a_1^\dagger a_2^\dagger-a_1a_2\right),
\end{eqnarray}
which is a two-mode squeezing Hamiltonian.
Note that Eq.~(\ref{eq_cdzz1}) does not yield beam-splitter coupling, because if we choose the approximation of $\sum_{s_1s_2}s_i|E_{s_1s_2}(t)\rangle\langle E_{s_1s_2}(t)|$ such that $a_1^\dagger a_2$ and $a_2^\dagger a_1$ appear, then $H_1(t)$ becomes zero.

\subsection{Approximate counter term for $R_x$}
For $R_x$, Eq.~(\ref{eq_variational}) gives $\beta(t)=\sqrt{[p+g_0(t)]/K}$, and the eigenstates can be approximated by
\begin{eqnarray}
	|E_0(t)\rangle&\simeq&\frac{1}{\sqrt{2}}\left[|\beta(t)\rangle+|\!-\!\beta(t)\rangle\right],\\
	|E_1(t)\rangle&\simeq&\frac{1}{\sqrt{2}}\left[|\beta(t)\rangle-|\!-\!\beta(t)\rangle\right],
\end{eqnarray}
where $|\langle\beta(t)|\!-\!\beta(t)\rangle|\ll1$ is assumed.
The counter term in Eq.~(\ref{eq_cd}) becomes 
\begin{eqnarray}
	H_1(t)&\simeq&\frac{i\dot{g}_0(t)}{4K\beta(t)}\left[a^\dagger Z(t)-Z(t)a\right],
\end{eqnarray}
where $Z(t)=|\beta(t)\rangle\langle\beta(t)|-|{\!-\!\beta(t)}\rangle\langle\!-\!\beta(t)|$ is a $Z$ operator in the qubit space.
Using $Z(t)=P(t)aP(t)/\beta(t)=P(t)a^\dagger P(t)/\beta(t)$, ignoring $P(t)$ similarly to $R_{zz}$, and assuming $|g_0(t)/p|\ll1$, we obtain nonzero Hermitian $H_1(t)$ as follows:
\begin{eqnarray}
	H_1(t)&\simeq&\frac{i\dot{g}_0(t)}{4K\alpha^2}\left(a^{\dagger2}-a^2\right).
\end{eqnarray}

\section{Properties of counter terms}
The exact counter term in Eq.~(\ref{eq_H1}) can be rewritten, by using time derivative of the eigenvalue equation in Eq.~(\ref{eq_eigen}), as~\cite{Berry2009, Guery-Odelin2019}
\begin{eqnarray}
		\!H_1(t)\!&=&\!i\dot{g}_0(t)\!\sum_{k\neq m}\!\frac{|E_k(t)\rangle\langle E_k(t)|A_0|E_m(t)\rangle\langle E_m(t)|}{E_m(t)-E_k(t)}\!,\!\nonumber\\
		\label{eq_H1E}
\end{eqnarray}
wehre $H_0(t)$ in Eqs.~(\ref{eq_H01bit}) and (\ref{eq_H02bit}) have been assumed.
\subsection{Matrix elements}\label{sec_Counter}
The matrix elements for $k\neq m$ are
\begin{eqnarray}
	\!\langle E_k(t)|H_1(t)|E_m(t)\rangle\!&=&\!i\dot{g}_0(t)\!\frac{\langle E_k(t)|A_0|E_m(t)\rangle}{E_m(t)-E_k(t)}\!.\!\label{eq_MatEle}
\end{eqnarray}
Equation~(\ref{eq_MatEle}) means that to cancel unwanted transitions, when the matrix elements of $A_0$ [and $\dot{g}_0(t)$] are nonzero, the corresponding matrix elements of a counter term $H_1(t)$ must be also nonzero.
Thus, an approximate $H_1(t)$ may be more effective when its matrix elements are more similar to those of $A_0$.
As mentioned in Sec.~\ref{sec_ApproxCounter}, we think that for the above reason, $H_1(t)$ with the two-mode squeezing in Eq.~(\ref{eq_cdzz2}) works better for $A_0$ with the two-mode squeezing in Eq.~(\ref{eq_TwoMode}) than $A_0$ with the beam-splitter coupling in Eq.~(\ref{eq_BeamSplitter}).

Also, to have nonzero matrix elements in common, $H_1(t)$ must have the same symmetries as $A_0$.
On the other hand, as mentioned in Sec.~\ref{sec_ApproxCounter}, the counter term with the beam-splitter coupling in Eq.~(\ref{eq_BeamSplitterOrth}) has permutation symmetry different from $A_0$ in Eqs.~(\ref{eq_BeamSplitter}) and (\ref{eq_TwoMode}), that is, the interchange of KPO1 and 2 leads to a sign change in Eq.~(\ref{eq_BeamSplitterOrth}) while does not in $A_0$.
Thus, the counter term in Eq.~(\ref{eq_BeamSplitterOrth}) may not be effective.

\subsection{Time reversal symmetry}\label{sec_TimeReversal}
Basd on Eq.~(\ref{eq_H1E}), we here consider the symmetry of $H_1(t)$ with respect to a time reversal $t\to T-t$.
When $g_0(t)$ and hence $H_0(t)$ are symmetric, namely, $H_0(T-t)=H_0(t)$, the eigenvalue equation in Eq.~(\ref{eq_eigen}) indicates that $E_k(t)$ and $|E_k(t)\rangle$ are also symmetric, that is, they can be chosen to be $E_k(T-t)=E_k(t)$ and $|E_k(T-t)\rangle=|E_k(t)\rangle$.
Also, when $g_0(t)$ is symmetric, $\dot{g}_0(t)$ is antisymmetric, $\dot{g}_0(T-t)=-\dot{g}_0(t)$.
Equation~(\ref{eq_H1E}) then indicates that symmetric $g_0(t)$ yields antisymmetric $H_1(t)$.
We thus use antisymmetric $g_1(t)$ in Eq.~(\ref{eq_g1}) for numerical optimization.

\subsection{One-parameter continuous gate with a counter term}\label{sec_ContinuousGate}
Equation~(\ref{eq_H1E}) indicates that a scaled $\lambda g_0(t)$ does not necessary scales $H_1(t)$ to $\lambda H_1(t)$, because $H_1(t)$ depends on $g_0(t)$ through $|E_k(t)\rangle$ and $E_k(t)$.
If the dependence of $|E_k(t)\rangle$ and $E_k(t)$ on $g_0(t)$ is negligible, such a scaling holds.
We think that this is an explanation for the one-parameter continuous gates with the counter terms for $R_z$ and $R_{zz}$ shown in Figs.~\ref{fig_ErPhi}(a) and \ref{fig_WaveRzz}(a).
Also, as mentioned in Sec.~\ref{sec_Rx}, since $R_x$ largely changes the state during the gate operation, the one-parameter continuous gate would not work.

\section{Average gate fidelities}\label{sec_AverageFidelity}
When dissipation is not included, we calculate a gate fidelity averaged over all initial states in a qubit space by~\cite{Pedersen2007}
\begin{eqnarray}
	\bar{F}=\frac{1}{d(d+1)}\left[\left|\mathrm{tr}\left(U_0^\dagger U\right)\right|^2+\mathrm{tr}\left(UU^\dagger\right)\right],\label{eq_BarF}
\end{eqnarray}
where $d=2, 4$ is the dimension of the qubit space for a single- and two-qubit gate, respectively, $U_0=R_z, R_x, R_{zz}$ is an ideal gate operation, and $U$ is a time-evolution operator projected onto the qubit space.
For a single-qubit gate, $U$ can be given by
\begin{eqnarray}
	U=\left(\begin{array}{cc}
		\langle\tilde{0}|\psi_0\rangle&\langle\tilde{0}|\psi_1\rangle\\
		\langle\tilde{1}|\psi_0\rangle&\langle\tilde{1}|\psi_1\rangle
	\end{array}\right),
\end{eqnarray}
where $|\psi_0\rangle, |\psi_1\rangle$ are states after time evolution for the gate time $T$ calculated with the Schr\"{o}dinger equation in Eq.~(\ref{eq_Schr}) with the initial states $|\tilde{0}\rangle, |\tilde{1}\rangle$, respectively.
$U$ for a two-qubit gate can be calculated similarly with the initial states $|\tilde{0}\tilde{0}\rangle, |\tilde{0}\tilde{1}\rangle, |\tilde{1}\tilde{0}\rangle, |\tilde{1}\tilde{1}\rangle$.

When the single-photon loss is included, we calculate the following average gate fidelity,
\begin{eqnarray}
	\bar{F}_{\rm loss}=\frac{1}{N_{\rm init}}\sum_{l=1}^{N_{\rm init}}\langle\psi_{\rm init}^{(l)}|U_0^\dagger\rho_lU_0|\psi_{\rm init}^{(l)}\rangle,\label{eq_TilBarF}
\end{eqnarray}
where $|\psi_{\rm init}^{(l)}\rangle$ is an initial state and $\rho_l$ is the density operator of a final state calculated from the initial state $|\psi_{\rm init}^{(l)}\rangle\langle\psi_{\rm init}^{(l)}|$ with the master equation in Eq.~(\ref{eq_master}).
$N_{\rm init}$ is the number of the initial states.
For a single-qubit gate, we choose the following six initial states for $l=1, 2, \cdots$, 6,
\begin{eqnarray}
	|\psi_{\rm init}^{(l)}\rangle&=&\left(\!\begin{array}{c}
			1\\
			0
		\end{array}\!\right)\!, \!\left(\!\begin{array}{c}
			0\\
			1
		\end{array}\!\right)\!, \!\frac{1}{\sqrt{2}}\!\left(\!\begin{array}{c}
			1\\
			1
		\end{array}\!\right)\!, \!\frac{1}{\sqrt{2}}\!\left(\!\begin{array}{c}
			1\\
			-1
		\end{array}\!\right)\!,\nonumber\\
		&&\!\frac{1}{\sqrt{2}}\!\left(\!\begin{array}{c}
			1\\
			i
		\end{array}\!\right)\!, \!\frac{1}{\sqrt{2}}\!\left(\!\begin{array}{c}
			1\\
			-i
		\end{array}\!\right)\!.
\end{eqnarray}
For a two-qubit gate, we use the 36 initial states given by $|\psi_{\rm init}^{(l)}\rangle|\psi_{\rm init}^{(l')}\rangle$ with $l, l'=1, 2,\cdots,6$.
We numerically find that $\bar{F}_{\rm loss}$ is in good agreement with $\bar{F}$ in the absence of the single-photon loss.

The average infidelity due to the single-photon loss alone can be approximated well by~\cite{Puri2017a, Aoki2023}
\begin{eqnarray}
	1-\bar{F}_{\rm loss}\simeq\frac{1}{3}\left(1-e^{-2\alpha^2\kappa T}\right),\label{eq_Floss}
\end{eqnarray}
per KPO for the large mean photon number $\alpha^2\gg1$.

\section{Waveforms of pulse amplitudes}\label{sec_wave}
\subsection{Waveforms for $R_z$}
\begin{enumerate}
	\item Analytic waveforms~\cite{Goto2016a} without the counter term:
		\begin{eqnarray}
			g_0(t)&=&\frac{\pi\phi}{8T\alpha}\sin\frac{\pi t}{T},\\
			g_1(t)&=&0.
		\end{eqnarray}
	\item Analytic waveforms with the counter term in Eq.~(\ref{eq_cdz}): 
		\begin{eqnarray}
			g_0(t)&=&\frac{\pi\phi}{8T\alpha}\sin\frac{\pi t}{T},\\
			g_1(t)&=&\frac{\dot{g}_0(t)}{4K\alpha^2}.\label{eq_g1aRz}
		\end{eqnarray}
	\item Numerically optimized waveforms in Eq.~(\ref{eq_g0}) without the counter term:
		Initial $g_{j,n}$ are
		\begin{eqnarray}
			g_{0,2}&=&\frac{\phi}{2T\alpha},\\
			g_{j,n}&=&0\text{ for the others}.
		\end{eqnarray}
	\item Numerically optimized waveforms in Eqs.~(\ref{eq_g0}) and (\ref{eq_g1}) with the counter term:
		Initial $g_{j,n}$ are, cf. Eq.~(\ref{eq_g1aRz}),
		\begin{eqnarray}
			g_{0,2}&=&\frac{\phi}{2T\alpha},\\
			g_{1,1}&=&\frac{\pi g_{0,2}}{4KT\alpha^2},\\
			g_{j,n}&=&0\text{ for the others}.
		\end{eqnarray}
\end{enumerate}

\subsection{Waveforms for $R_{zz}$}
\begin{enumerate}
	\item Analytic waveforms~\cite{Goto2016a} without the counter term:
		\begin{eqnarray}
			g_0(t)&=&\frac{\pi\Theta}{8T\alpha^2}\sin\frac{\pi t}{T},\\
			g_1(t)&=&0.
		\end{eqnarray}
	\item Analytic waveforms with the counter term in Eq.~(\ref{eq_cdzz2}): 
		\begin{eqnarray}
			g_0(t)&=&\frac{\pi\Theta}{8T\alpha^2}\sin\frac{\pi t}{T},\\
			g_1(t)&=&\frac{\dot{g}_0(t)}{2K\alpha^2}.\label{eq_g1aRzz}
		\end{eqnarray}
	\item Numerically optimized waveforms in Eq.~(\ref{eq_g0}) without the counter term:
		Initial $g_{j,n}$ are
		\begin{eqnarray}
			g_{0,2}&=&\frac{\Theta}{2T\alpha^2},\\
			g_{j,n}&=&0\text{ for the others}.
		\end{eqnarray}
	\item Numerically optimized waveforms in Eqs.~(\ref{eq_g0}) and (\ref{eq_g1}) with the counter term:
		Initial $g_{j,n}$ are, cf. Eq.~(\ref{eq_g1aRzz}),
		\begin{eqnarray}
			g_{0,2}&=&\frac{\Theta}{2T\alpha^2},\\
			g_{1,1}&=&\frac{\pi g_{0,2}}{2KT\alpha^2},\\
			g_{j,n}&=&0\text{ for the others}.
		\end{eqnarray}
\end{enumerate}

\subsection{Waveforms for $R_x$}
\begin{enumerate}
	\item Analytic waveforms~\cite{Goto2016a} without the counter term:
		The following $\Delta$ is determined by maximizing $\bar{F}$.
		\begin{eqnarray}
			g_0(t)&=&\frac{\Delta}{2}\left(1-\cos\frac{2\pi t}{T}\right),\label{eq_g0Rx}\\
			g_1(t)&=&0.
		\end{eqnarray}
	\item Analytic waveforms with the counter term in Eq.~(\ref{eq_cdx}):
		The following $\Delta_{\rm count}$ is determined as above.
		\begin{eqnarray}
			g_0(t)&=&\frac{\Delta_{\rm count}}{2}\left(1-\cos\frac{2\pi t}{T}\right),\label{eq_g0RxSTA}\\
			g_1(t)&=&\frac{\dot{g}_0(t)}{4K\alpha^2}.\label{eq_g1aRx}
		\end{eqnarray}
	\item Numerically optimized waveforms in Eq.~(\ref{eq_g0}) without the counter term:
		With $\Delta$ determined for the analytic waveforms in Eq.~(\ref{eq_g0Rx}), initial $g_{j,n}$ are
		\begin{eqnarray}
			g_{0,2}&=&\Delta,\\
			g_{j,n}&=&0\text{ for the others}.
		\end{eqnarray}
	\item Numerically optimized waveforms in Eqs.~(\ref{eq_g0}) and (\ref{eq_g1}) with the counter term:
		With $\Delta_{\rm count}$ determined for the analytic waveforms in Eq.~(\ref{eq_g0RxSTA}), initial $g_{j,n}$ are, cf. Eq.~(\ref{eq_g1aRx}),
		\begin{eqnarray}
			g_{0,2}&=&\Delta_{\rm count},\\
			g_{1,1}&=&\frac{\pi g_{0,2}}{4KT\alpha^2},\\
			g_{j,n}&=&0\text{ for the others}.
		\end{eqnarray}
\end{enumerate}

\section{The Wigner function during $R_x$}\label{sec_WignerRx}
\begin{figure}
	\includegraphics[width=8.5cm]{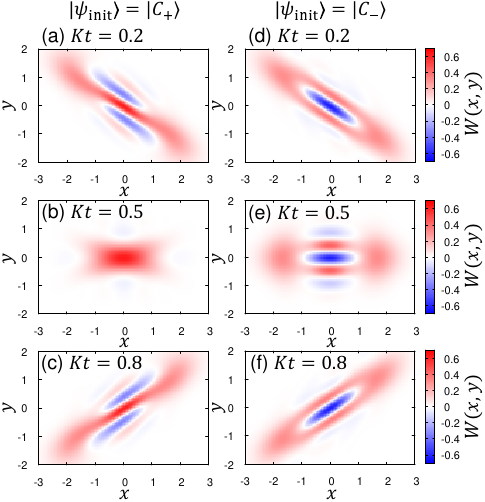}
	\caption{The Wigner functions during time evolution under the optimized $g_j(t)$ in Fig.~\ref{fig_WaveRx}(a) with the initial states (a)-(c) $|C_+\rangle$ and (d)-(f) $|C_-\rangle$.
		\label{fig_WignerRx}}
\end{figure}
Figure~\ref{fig_WignerRx} shows the Wigner function during the optimized $R_x$ in Fig.~\ref{fig_WaveRx}(a).
Figure~\ref{fig_WignerRx}(b) shows that for $|\psi_{\rm init}\rangle=|C_+\rangle$ the intermediate state looks like a vacuum state, which agrees with the small mean photon number in Fig.~\ref{fig_WaveRx}(b).
The vacuum state may be realized because for large $|g_0(t)|/K$, $H_0(t)$ can be approximated by $g_0(t)a^\dagger a$ and the vacuum state becomes an eigenstate.
On the other hand, Fig.~\ref{fig_WignerRx}(e) shows that for $|\psi_{\rm init}\rangle=|C_-\rangle$ the intermediate state resembles $|C_-\rangle$, which is consistent with the large population in the qubit space for $Kt=0.5$ in Fig.~\ref{fig_WaveRx}(c).


\end{document}